\documentclass[nohyperref]{article}

\usepackage{microtype}
\usepackage{booktabs} 

\usepackage{hyperref}
\usepackage{graphicx}
\usepackage{subfigure}
\usepackage{caption}
\usepackage{comment}

\usepackage[accepted]{icml2022}

\usepackage{amsmath}
\usepackage{amssymb}
\usepackage{mathtools}
\usepackage{amsthm}

\usepackage[capitalize,noabbrev]{cleveref}

\theoremstyle{plain}

\theoremstyle{definition}

\theoremstyle{remark}

\usepackage[textsize=tiny]{todonotes}

\icmltitlerunning{Outlier Detection using Self-Organizing Maps for Automated Blood Cell Analysis}

\begin{document}

\twocolumn[
\icmltitle{Outlier Detection using Self-Organizing Maps\\for Automated Blood Cell Analysis}



\icmlsetsymbol{equal}{*}

\begin{icmlauthorlist}
\icmlauthor{Stefan R\"ohrl}{equal,ldv}
\icmlauthor{Alice Hein}{equal,ldv}
\icmlauthor{Lucie Huang}{equal,ldv}
\icmlauthor{Dominik Heim}{lbe}
\icmlauthor{Christian Klenk}{lbe}
\icmlauthor{Manuel Lengl}{ldv}
\icmlauthor{Martin Knopp}{ldv,lbe}
\icmlauthor{Nawal Hafez}{ldv}
\icmlauthor{Oliver Hayden}{lbe}
\icmlauthor{Klaus Diepold}{ldv}
\end{icmlauthorlist}

\icmlaffiliation{ldv}{Chair of Data Processing, Technical University of Munich, Germany}
\icmlaffiliation{lbe}{Heinz-Nixdorf Chair of Biomedical Electronics, Technical University of Munich, Germany}

\icmlcorrespondingauthor{Stefan R\"ohrl}{stefan.roehrl@tum.de}

\icmlkeywords{Machine Learning, Outlier Detection, Self-Organizing Maps, Digital Holographic Microscopy, Quantitative Phase Imaging, Blood Cell Analysis}

\vskip 0.3in
]



\printAffiliationsAndNotice{\icmlEqualContribution} 

\begin{abstract}
The quality of datasets plays a crucial role in the successful training and deployment of deep learning models. Especially in the medical field, where system performance may impact the health of patients, clean datasets are a safety requirement for reliable predictions. Therefore, outlier detection is an essential process when building autonomous clinical decision systems. In this work, we assess the suitability of Self-Organizing Maps for outlier detection specifically on a medical dataset containing quantitative phase images of white blood cells. We detect and evaluate outliers based on quantization errors and distance maps. Our findings confirm the suitability of Self-Organizing Maps for unsupervised Out-Of-Distribution detection on the dataset at hand. Self-Organizing Maps perform on par with a manually specified filter based on expert domain knowledge. Additionally, they show promise as a tool in the exploration and cleaning of medical datasets. As a direction for future research, we suggest a combination of Self-Organizing Maps and feature extraction based on deep learning.
\end{abstract}

\section{Introduction}
\label{intro}
Nowadays, many diseases like leukemia are diagnosed by analyzing blood samples and detecting unhealthy distributions of different types of blood cells \cite{mittal2022automated}. Therefore, analysis of cellular structures make up a large part of medical laboratory tests. However, currently used gold standards of hematological analysis either have the disadvantage that they cannot classify certain cell types or are associated with a high manual effort \cite{meintker2013comparison, Filby2016}. Computer vision and machine learning (ML) in combination with contrast-rich digital holographic microscopy has the potential to perform such hematological analyses in a more cost effective, flexible and faster way \cite{Jo2018}. 

Unfortunately, during the process of data collection, outliers such as defocused cells, duplets and debris may occur due to activation, apoptosis, and aggregation of cells or insufficient flow focusing. In the training stage, including these outliers in one's dataset may deteriorate model performance, since there is also no industrial grade calibrator for this holographic flow cytometry assay. In a production environment, outliers may even pose a safety issue if the model cannot reliably recognize them as such, potentially leading to a wrong classification of, say, debris as an interesting event. 
In this work, we examine the suitability of Self-Organizing Maps (SOMs) as a tool for the detection of outliers in a dataset of holographic microscopic images of white blood cells (WBCs). We first provide some background on SOMs in Section \ref{background} and describe our dataset and experimental setup in Section \ref{methods}. Section \ref{results} presents our results. We end with a brief discussion of related work in Section \ref{relatedwork} and ways our approach could be expanded upon in Section \ref{conclusion}.

\section{Background}
\label{background}
The SOM is an unsupervised artificial neural network first proposed by Teuvo Kohonen (\citeyear{kohonen1990self}) in early 1981. This dimensionality reduction technique groups data points into clusters on a 2D lattice according to their mutual similarity. The lattice space of a SOM consists of a predefined number of neurons. Each neuron has its own weight vector, which is initialized through some initialization function (e.g. principal component analysis). The weight vector of a neuron $j$ can be described as
\begin{equation}
    w_{j} = [w_{j1}, w_{j2}, ..., w_{jd}]^{T},~j=1,2,...,J , \nonumber
\end{equation}
where $J$ is the number of neurons and $d$ the number of input features.

The SOM is then trained for a set number of iterations by choosing an input data point $x \in \mathbb{R}^d$ from the training dataset and computing its activation distance to all other neurons. The index $c = c(x)$ of the neuron with the closest Euclidean distance to $x$, also called the Best Matching Unit (BMU), is determined using 
\begin{equation}
    c(x) = \underset{j}{arg min} \left \| x - w_{j} \right \|,~j=1,2,...,J. \nonumber
\end{equation}
Based on a predefined spread (e.g. standard deviation $\sigma$), a neighborhood kernel $h_{j,c(x)}(n)$ controls the update influence on the surrounding of the BMU. The weight vectors of the BMU and its neighbors $w_{j}$ are then updated according to a time-variant learning rate $\alpha(n)$ using
\begin{equation}
    w_{j}(n+1) = w_{j}(n)+\alpha(n) \cdot h_{j,c(x)}(n) \cdot (x(n)-w_{j}(n)), \nonumber
\end{equation}
where $n$ stands for the current iteration. Algorithm \ref{algo:som_training} summarizes this process.

After successful training, the weight vectors of the SOM have adjusted to reflect the distribution of the input data in a topology-preserving manner: data points which are similar to each other in the input space are matched onto neurons close to each other in the lattice space \cite{kaski1997data}. This is a useful property for the detection of outliers within a large dataset, as inliers are expected to form large and dense clusters of neurons in the lattice space. Outliers on the other hand are expected to be scattered across the lattice space with a large distance from the dense clusters. 

\begin{algorithm}
\begin{algorithmic}[1]
\State initialize weight vectors $w$ of all neurons
\State $N \gets$ \emph{number of iterations}
\State $J \gets$ \emph{number of neurons}
\For{$n \gets 1$ to $N$}
	\State $x \gets$ random input data point from the input dataset
	\For{$j \gets 1$ to $J$}
		\State calculate distance $d_{j}(n) = \left \| x - w_{j}(n) \right \|$ 
	\EndFor
	\State calculate index for BMU $c(x) = \underset{j}{arg min}~d_j(n)$
	\State determine neighborhood function $h_{j,c(x)}(n)$ based on $\sigma$ and $c(x)$
	\For{$j \gets 1$ to $J$}
		\State update weights with $w_{j}(n+1)$ \par 
		\hskip\algorithmicindent$= w_{j}(n)+\alpha(n) \cdot h_{j,c(x)}(n) \cdot (x(n)-w_{j}(n))$
	\EndFor
\EndFor
\end{algorithmic}
\caption{SOM training algorithm}
\label{algo:som_training}
\end{algorithm}

\section{Methods}
\label{methods}

\subsection{Data}
\label{dataset}
The dataset used in this work\footnote{All human samples were collected with informed consent and procedures approved by application 620/21 S-KK of the ethic committee of the Technical University of Munich.} consists of quantitative phase images of four types of WBCs (eosinophils, lymphocytes, monocytes, and neutrophils), taken by a digital holographic microscope. These images of size 512$\times$384 pixels contain multiple cells per image and represent their optical density. Using threshold segmentation, the raw phase images are segmented to yield single cell image patches of size 50$\times$50 pixels. Examples can be seen in Figure \ref{fig:examples}. For this work, we used three segmented datasets:
\begin{itemize}
	\item \textbf{Unfiltered dataset}, 447,541 images \newline 
	This dataset contains images of 41,881 eosinophils, 77,672 lymphocytes, 58,760 monocytes and 269,228 neutrophils. Since it has not been manually cleaned, there are an unknown number of inliers and outliers. 

	\item \textbf{Inlier dataset}, 82,056 images \newline
	This dataset was created by filtering images based on predefined thresholds for the four morphological features \textit{optical height max}, \textit{circularity}, \textit{area} and \textit{equivalent diameter} of each cell. The four classes of WBCs are balanced to 20,514 images per class.
	
	\item \textbf{Outlier dataset}, 10,136 images \newline
	The dataset contains 352 images captured with focus set 7.5 $\mu m$ over the ideal focus and 803 images with focus set 15 $\mu m$ over the ideal focus. 7,749 images contain high background noise and 1,232 images were captured at the border of the microfluidic channel, which leads to high interferences due to light scattering. 
\end{itemize}

All segmented images were normalized to the range of the inlier dataset, and six ($d=6$) morphological features were extracted, namely \textit{area}, \textit{circularity},\textit{ equivalent diameter}, \textit{optical height max}, \textit{optical height variance}, and \textit{energy} \cite{ugele2018labelwiley, roehrl2019}.

\subsection{Experiments}
\label{experiments}
After preprocessing, we trained a SOM on the inlier dataset, then tested the model on the outlier and unfiltered dataset and evaluated the detected outliers and inliers. For evaluation, we used the \emph{average quantization error}, which is the normed average of the quantization errors of all input samples, calculated using
\begin{equation}
    E_{AQ} = \frac{1}{M} \sum_{i=1}^{M} \left \| x_{i}-w_{c} \right \| ~~\text{with}~c = \underset{j}{arg min} \left \| x_{i} - w_{j} \right \|. \nonumber
\end{equation}
Here, $M$ is the size of the input dataset and $j$ the index of the respective neuron. The smaller the average quantization error, the better a fixed-sized SOM reflects the input dataset. We defined samples with a quantization error greater than the $2\sigma$ deviation of all quantization errors as outliers. 

As per Kohonen's recommendation (\citeyear{kohonen1990self}), our SOM consisted of $5\sqrt{K}$ neurons, where $K$ is the cardinality of our inlier dataset. Its shape was chosen such that its ratio of height to width equaled the ratio of the two largest eigenvalues of its autocorrelation matrix \cite{ponmalai2019self}, resulting in a 65$\times$22 lattice. The SOM was trained with a sigma of $1$, learning rate of $1$, hexagonal topology, gaussian neighborhood function and euclidean activation distance, as this was found to be a suitable hyperparameter configuration in preliminary tests with a 5-fold cross-validation, leading to the lowest quantization error. All experiments were implemented in Python and made use of the Scikit-learn\footnote{\scriptsize\url{https://scikit-learn.org}}, OpenCV\footnote{\scriptsize\url{https://opencv.org}}, TensorFlow\footnote{\scriptsize\url{https://tensorflow.org}}, Keras\footnote{\scriptsize\url{https://keras.io}}, and MiniSOM\footnote{\scriptsize\url{https://github.com/JustGlowing/minisom}} libraries.

\section{Results}
\label{results}
\begin{figure}
    \centering
    \includegraphics[width=0.48\textwidth]{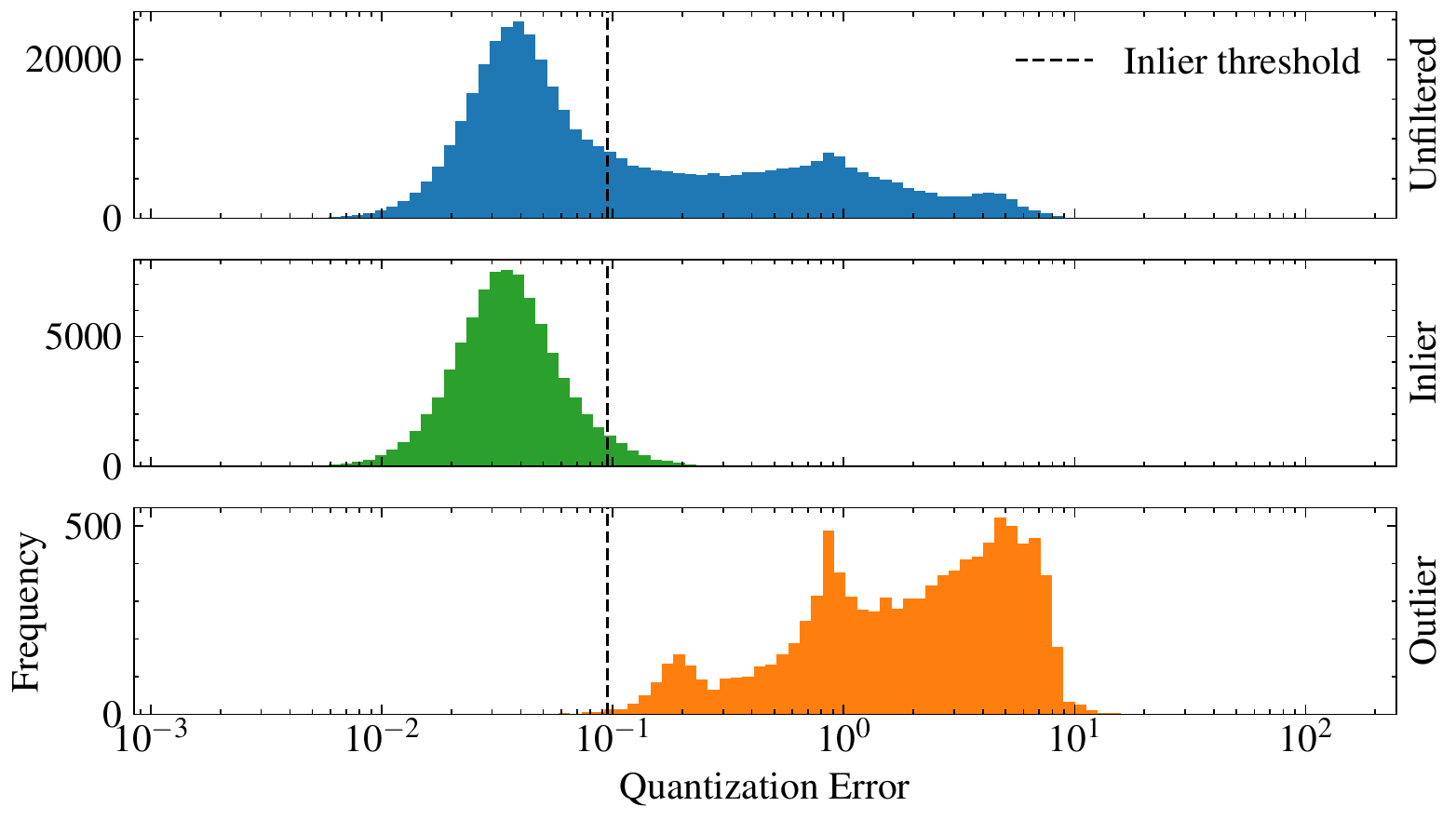}
    \caption{Quantization error distributions for the three datasets}
    \label{fig:hist}
\end{figure}
\begin{figure}[h]
\centering
\subfigure[Quantization error range 0.0 -- 0.1]{\label{fig:range0}\includegraphics[width=80mm]{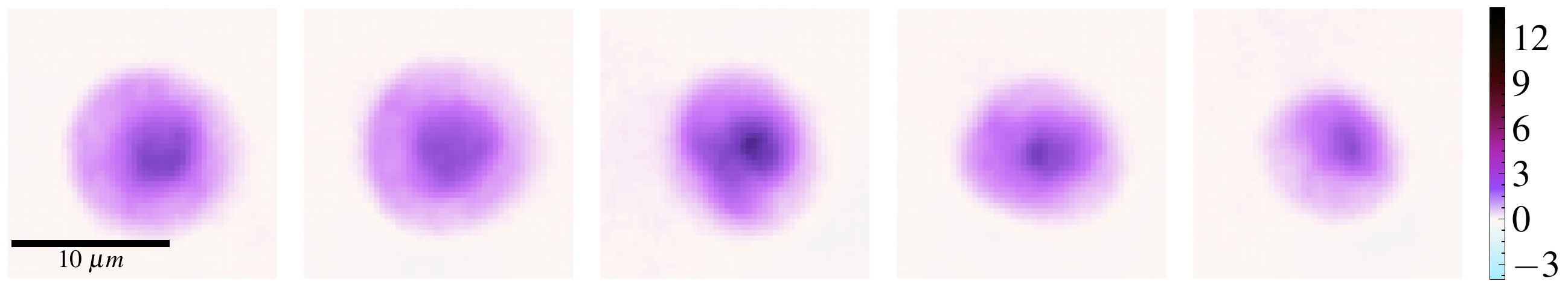}}
\subfigure[Quantization error range 0.5 -- 0.6]{\label{fig:range05}\includegraphics[width=80mm]{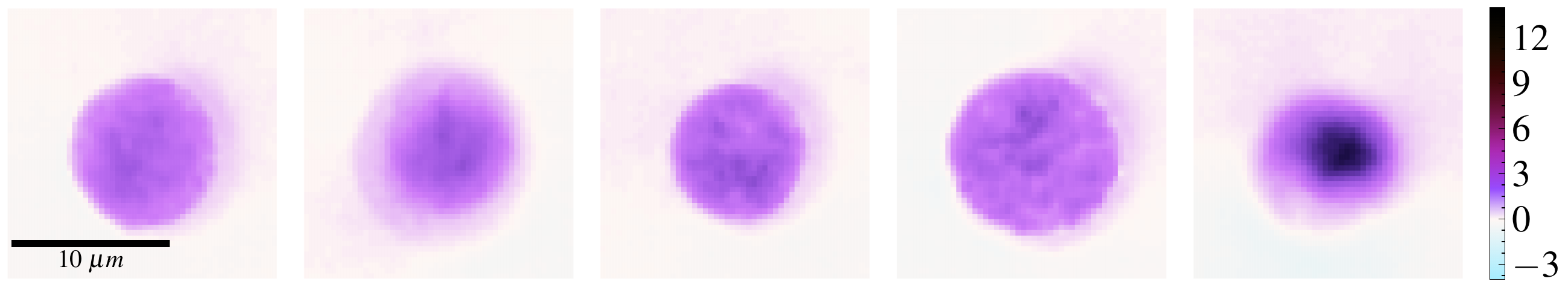}}
\subfigure[Quantization error range 1.0 -- 2.0]{\label{fig:range1}\includegraphics[width=80mm]{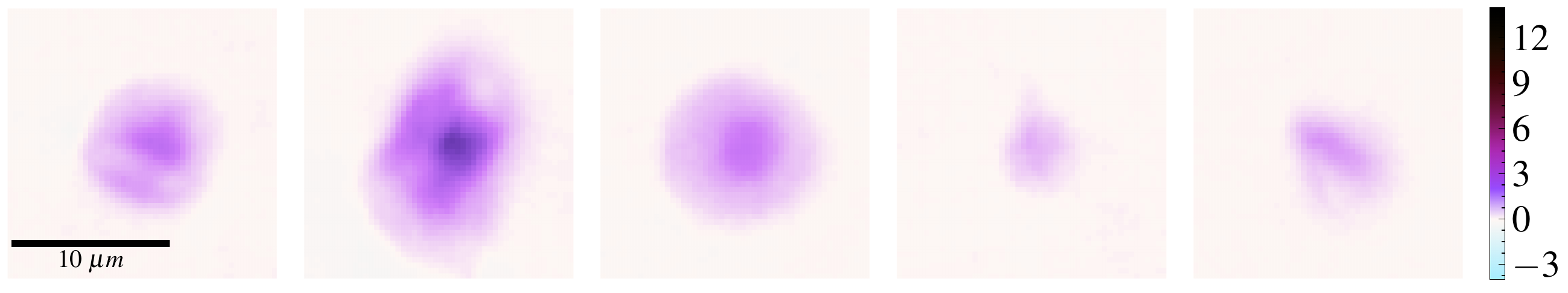}}
\subfigure[Quantization error range 3.0 -- 4.0]{\label{fig:range3}\includegraphics[width=80mm]{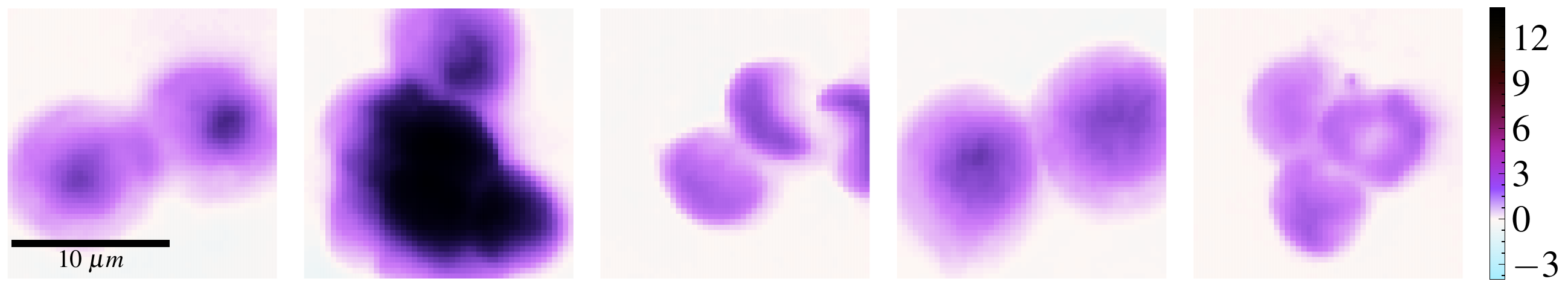}}
\subfigure[Quantization error range 10.0 -- 20.0]{\label{fig:range10}\includegraphics[width=80mm]{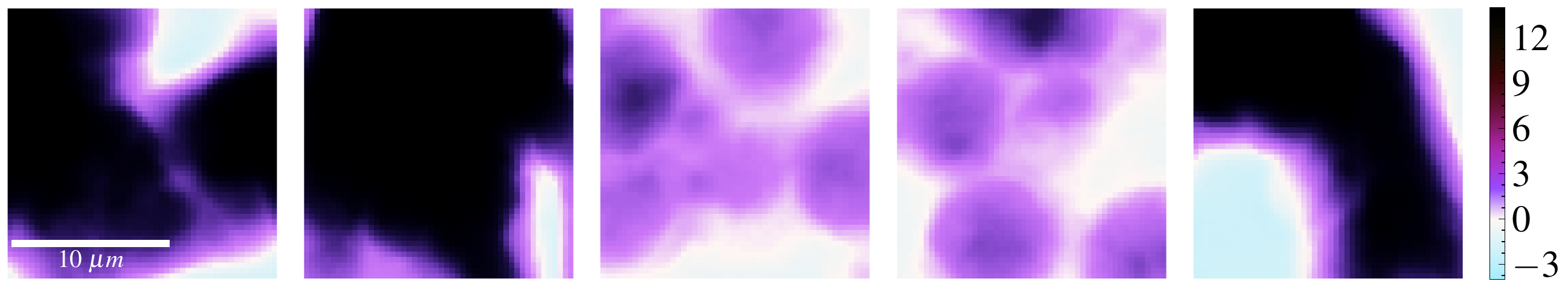}}
\subfigure[Quantization error range 30.0 -- 100.0]{\label{fig:range30}\includegraphics[width=80mm]{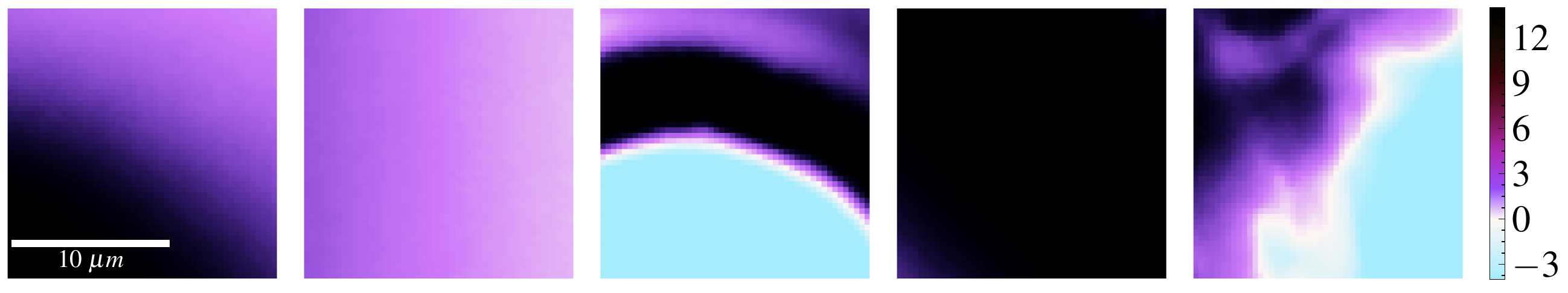}}
\caption{Examples of inliers and outliers detected by the SOM in the unfiltered dataset}
\label{fig:examples}
\end{figure}

The middle graph of Figure \ref{fig:hist} shows the distribution of all inlier quantization errors. As can be seen, most of the errors fall within a small range around 0.04, which indicates that the SOM was trained to fit the inlier dataset well. According examples for inliers are displayed in Figure \ref{fig:range0}. Next, we evaluate the quantization errors of the outlier dataset. If the SOM worked perfectly, all errors should be greater than the inlier threshold. 

This is confirmed by Figure \ref{fig:hist} (bottom), where 99.6\% of all data are correctly detected as outliers. Finally, we tested the SOM on the unfiltered dataset consisting of an unknown number of unlabeled inliers and outliers. As expected, most of the quantization errors for the unfiltered dataset lay within the threshold of 0.095, while the rest stretches out to large quantization error ranges, yielding an outlier percentage of 43.26\%. That is approximately the same amount as detected with the currently used filtering method, which relies on manually specified feature thresholds based on domain expertise.

Taking a look at the inliers and outliers detected in the unfiltered dataset, we observe that in error range [0.5, 0.6], the detected outliers start to take on irregular shapes, such as too small or unclean circles. Cells in error range [1.0, 2.0] often have blurred and irregular contours. Range [3.0, 4.0] covers the case of double cells, which where mistaken for single cells in the segmentation process. Larger error ranges contain completely irregular cells or edge cases like the border of the microfluidic channel or air bubbles.

\begin{figure}
    \centering
    \includegraphics[height=39mm]{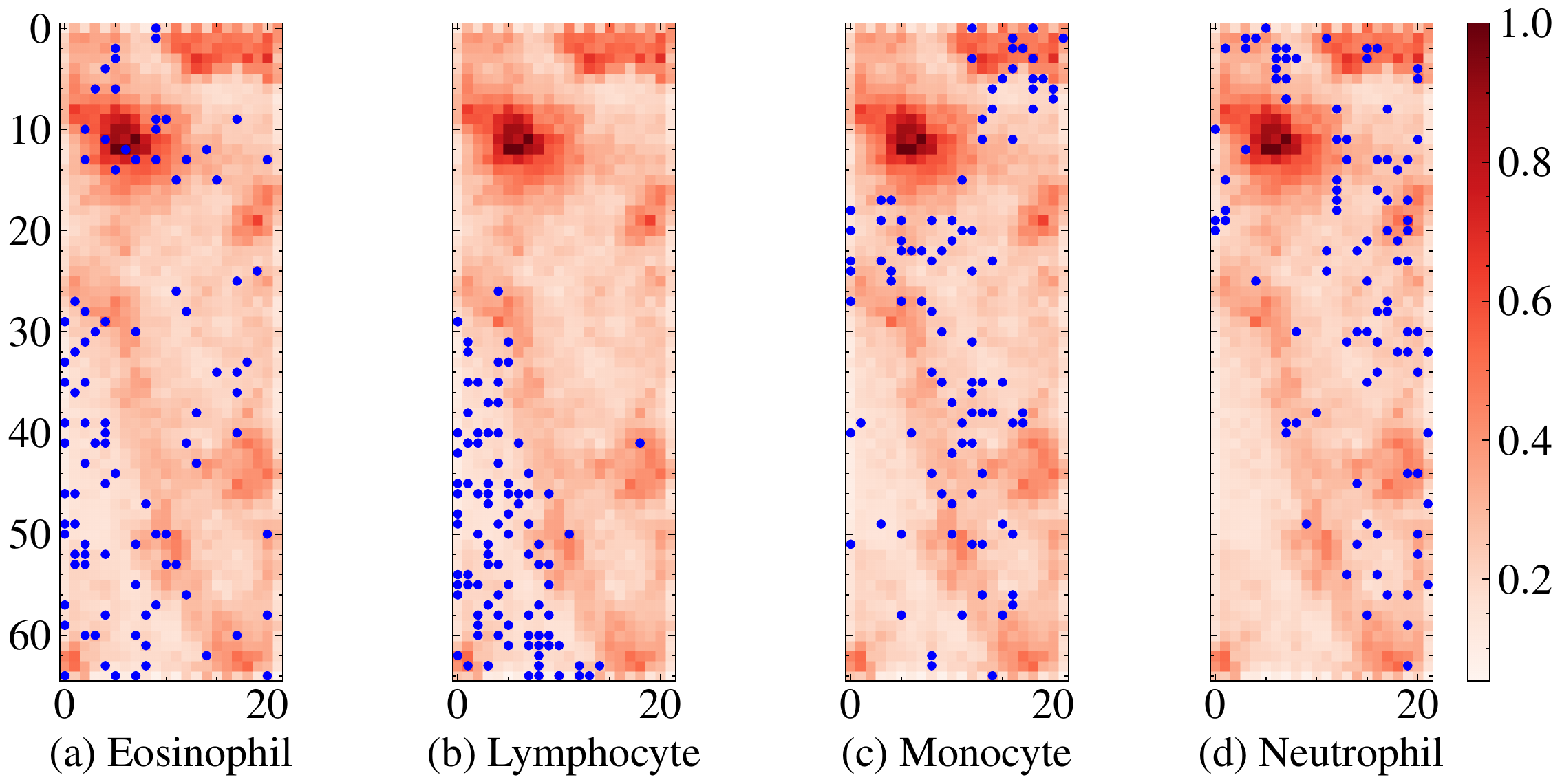}
    \caption{Positions of four inlier classes (a) eosinophil, (b) lymphocyte, (c) monocyte and (d) neutrophil on SOM distance map\label{fig:distance_map_inlier_classes}}
\end{figure}

A further evaluation technique we used was to inspect where on the SOM distance map the inliers and outliers were positioned. A distance map shows the distance of each neuron to its closest neighbors. The lighter the neuron, the smaller the distance to its neighbor neurons. Figure \ref{fig:distance_map_inlier_classes} displays the distance map as the aforementioned 65$\times$22 lattice as a background pattern. Each sub-figure shows that almost all inliers were plotted in light regions of the distance map, confirming the assumption that clusters with many neurons close to each other represent dense inlier classes. Additionally, the winning neurons of input data points from the same white blood cell classes formed clusters, suggesting that the SOM had not only learned to distinguish inliers and outliers, but also to some extent the four different classes of inliers. 

\begin{figure}
    \centering
    \includegraphics[height=39mm]{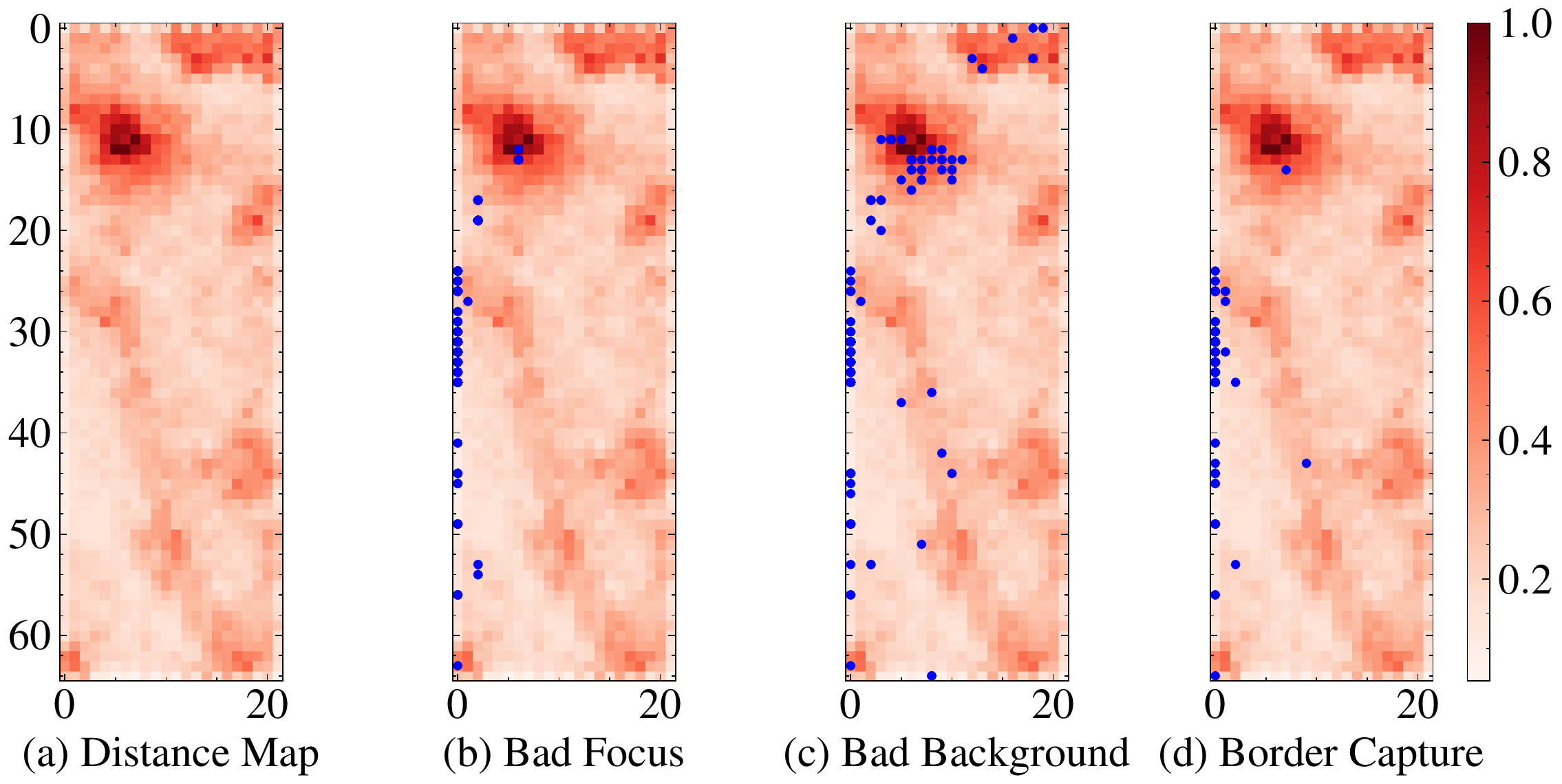}
    \caption{SOM distance map (a) and positions of outlier types (b) bad focus, (c) bad background and (d) border capture\label{fig:distance_map_bad}}
\end{figure}

This pattern is also confirmed by Figure \ref{fig:distance_map_bad}, which plots the positions of different types of outliers on the distance map. In contrast to the inlier data points, outliers tend to be positioned in darker, that is, less dense regions, or at the edge of the SOM. 

\section{Related Work}
\label{relatedwork}
Out-Of-Distribution (OOD) detection methods provide important safety mechanisms to prevent real-world systems from failing when confronted with anomalous data and have thus been the focus of much research. The three main categories of OOD detection approaches are classification-based, nearest neighbor-based, and clustering-based techniques \cite{chandola2009anomaly}, where SOMs can be said to belong to the latter category. Previous applications of SOMs for cluster-based OOD identification include intrusion detection \cite{labib2002nsom}, fault detection \cite{emamian2000robust} and fraud detection \cite{brockett1998using}. While this work uses a low-dimensional feature representation of the input objects, it is also possible to apply SOMs directly on pixel values as shown by \citet{Penn2002} on hyperspectral imagery data. \citet{Ruliang2018} extend this idea and combine the SOM with a deep neural network to obtain a \emph{change graph} in synthetic aperture radar images used for environmental monitoring. In the domain of tissue cell analysis \textit{in silico}, \citet{Yuan2021} presented a SOM for segmentation and classification. \citet{Rahmat2018} successfully demonstrate the morphological analysis of red blood cells which encourages the adaption of SOMs to WBCs in this work.

\section{Discussion and Conclusion}
\label{conclusion}

In this work, we confirmed the suitability of a SOM-based OOD detection approach on a dataset of holographic blood cell images. The SOM reached an accuracy of 99.69\% on a test set of outliers created through physical manipulations during the imaging or the sample preparation. When applied on a dataset with an unknown number of inliers and outliers, it performed similarly to a filter based on manually specified feature thresholds. Therefore, it spares the medical experts time consuming and expensive manual labor. The SOM-based method also enabled the observation of different types of outliers in different ranges of quantization errors, such as duplets and edge cases. This was not possible using the current filtering method. Hence, we achieved a more generalizable and robust approach to clean the vast holographic flow cytometry datasets. In addition, the optimized SOM could be used to distinguish between different classes of inliers, visible as separate clusters on the distance map.

However, the SOM still relies on extensive pre-processing to extract selected features. A next step would therefore be to take advantage of recent advances in deep learning by combining convolutional neural networks for feature extraction with SOMs for dimensionality reduction and OOD detection. Given the SOM’s clustering abilities, we also envision further applications such as dataset exploration and efficient data annotation by labelling entire parts of the SOM rather than individual examples.

\paragraph{Acknowledgement} The authors would like to especially honor the contributions of D. Heim for the sample preparation and recording of the measurements and L. Huang for the software implementation and experiments. This research was funded by the German Federal Ministry for Education and Research (BMBF) with the funding ID ZN 01 \textbar ~ S17049.


\bibliography{bibliography}

\begin{thebibliography}{17}
\providecommand{\natexlab}[1]{#1}
\providecommand{\url}[1]{\texttt{#1}}
\expandafter\ifx\csname urlstyle\endcsname\relax
  \providecommand{\doi}[1]{doi: #1}\else
  \providecommand{\doi}{doi: \begingroup \urlstyle{rm}\Url}\fi

\bibitem[Brockett et~al.(1998)Brockett, Xia, and Derrig]{brockett1998using}
Brockett, P.~L., Xia, X., and Derrig, R.~A.
\newblock {Using Kohonen’s Self-Organizing Feature Map to Uncover Automobile
  Bodily Injury Claims Fraud}.
\newblock \emph{Journal of Risk and Insurance}, 65\penalty0 (2):\penalty0
  245--274, 1998.

\bibitem[Chandola et~al.(2009)Chandola, Banerjee, and
  Kumar]{chandola2009anomaly}
Chandola, V., Banerjee, A., and Kumar, V.
\newblock {Anomaly detection: A survey}.
\newblock \emph{ACM computing surveys (CSUR)}, 41\penalty0 (3):\penalty0 1--58,
  2009.

\bibitem[Emamian et~al.(2000)Emamian, Kaveh, and Tewfik]{emamian2000robust}
Emamian, V., Kaveh, M., and Tewfik, A.~H.
\newblock {Robust Clustering of Acoustic Emission Signals Using the Kohonen
  Network}.
\newblock In \emph{Proceedings of the 2000 IEEE International Conference on
  Acoustics, Speech, and Signal Processing}, volume~6, pp.\  3891--3894. IEEE
  Computer Society, 2000.

\bibitem[Filby(2016)]{Filby2016}
Filby, A.
\newblock {Sample preparation for flow cytometry benefits from some lateral
  thinking}.
\newblock \emph{Cytometry Part A: the journal of the International Society for
  Analytical Cytology}, 89\penalty0 (12):\penalty0 1054--1056, 2016.

\bibitem[Jo et~al.(2018)Jo, Cho, Yun~Lee, Choi, Kim, Min, and Park]{Jo2018}
Jo, Y., Cho, H., Yun~Lee, S., Choi, G., Kim, G., Min, H.-s., and Park, Y.
\newblock {Quantitative Phase Imaging and Artificial Intelligence: A Review}.
\newblock \emph{IEEE Journal of Selected Topics in Quantum Electronics},
  25\penalty0 (1):\penalty0 1--14, 2018.

\bibitem[Kaski(1997)]{kaski1997data}
Kaski, S.
\newblock \emph{{Data Exploration Using Self-organizing Maps}}.
\newblock Acta polytechnica Scandinavica. Finnish Academy of Technology, 1997.

\bibitem[Kohonen(1990)]{kohonen1990self}
Kohonen, T.
\newblock The self-organizing map.
\newblock \emph{Proceedings of the IEEE}, 78\penalty0 (9):\penalty0 1464--1480,
  1990.

\bibitem[Labib \& Vemuri(2002)Labib and Vemuri]{labib2002nsom}
Labib, K. and Vemuri, R.
\newblock {NSOM: A real-time network-based intrusion detection system using
  self-organizing maps}.
\newblock \emph{Networks and Security}, 21\penalty0 (1), 2002.

\bibitem[Meintker et~al.(2013)Meintker, Ringwald, Rauh, and
  Krause]{meintker2013comparison}
Meintker, L., Ringwald, J., Rauh, M., and Krause, S.~W.
\newblock {Comparison of automated differential blood cell counts from Abbott
  Sapphire, Siemens Advia 120, Beckman Coulter DxH 800, and Sysmex XE-2100 in
  normal and pathologic samples}.
\newblock \emph{American journal of clinical pathology}, 139\penalty0
  (5):\penalty0 641--650, 2013.

\bibitem[Mittal et~al.(2022)Mittal, Dhalla, Gupta, and
  Gupta]{mittal2022automated}
Mittal, A., Dhalla, S., Gupta, S., and Gupta, A.
\newblock Automated analysis of blood smear images for leukemia detection: a
  comprehensive review.
\newblock \emph{ACM Computing Surveys (CSUR)}, 2022.

\bibitem[Penn(2002)]{Penn2002}
Penn, B.
\newblock Using self-organizing maps for anomaly detection in hyperspectral
  imagery.
\newblock In \emph{Proceedings, IEEE Aerospace Conference}, volume~3, pp.\
  1531--1535, 2002.

\bibitem[Ponmalai \& Kamath(2019)Ponmalai and Kamath]{ponmalai2019self}
Ponmalai, R. and Kamath, C.
\newblock {Self-Organizing Maps and Their Applications to Data Analysis}.
\newblock Technical report, Lawrence Livermore National Lab. (LLNL), Livermore,
  CA (United States), 2019.

\bibitem[Rahmat et~al.(2018)Rahmat, Wulandari, Faza, Muchtar, and
  Siregar]{Rahmat2018}
Rahmat, R.~F., Wulandari, F.~S., Faza, S., Muchtar, M.~A., and Siregar, I.
\newblock The morphological classification of normal and abnormal red blood
  cell using self organizing map.
\newblock \emph{{IOP} Conference Series: Materials Science and Engineering},
  308:\penalty0 012015, 2018.

\bibitem[R\"ohrl et~al.(2019)R\"ohrl, Ugele, Klenk, Heim, Hayden, and
  Diepold]{roehrl2019}
R\"ohrl, S., Ugele, M., Klenk, C., Heim, D., Hayden, O., and Diepold, K.
\newblock {Autoencoder Features for Differentiation of Leukocytes based on
  Digital Holographic Microscopy (DHM)}.
\newblock In \emph{Computer Aided Systems Theory - EUROCAST}, pp.\  281--288,
  2019.

\bibitem[Ugele et~al.(2018)Ugele, Weniger, Stanzel, Bassler, Krause, Friedrich,
  Hayden, and Richter]{ugele2018labelwiley}
Ugele, M., Weniger, M., Stanzel, M., Bassler, M., Krause, S.~W., Friedrich, O.,
  Hayden, O., and Richter, L.
\newblock Label-free high-throughput leukemia detection by holographic
  microscopy.
\newblock \emph{Advanced Science}, 5\penalty0 (12):\penalty0 1800761, 2018.

\bibitem[Xiao et~al.(2018)Xiao, Cui, Lin, Chen, Ni, and Lin]{Ruliang2018}
Xiao, R., Cui, R., Lin, M., Chen, L., Ni, Y., and Lin, X.
\newblock {SOMDNCD: Image Change Detection Based on Self-Organizing Maps and
  Deep Neural Networks}.
\newblock \emph{IEEE Access}, 6:\penalty0 35915--35925, 2018.

\bibitem[Yuan et~al.(2021)Yuan, Matusiak, Sirinukunwattana, Varma, Kidziński,
  and West]{Yuan2021}
Yuan, E., Matusiak, M., Sirinukunwattana, K., Varma, S., Kidziński, L., and
  West, R.
\newblock Self-organizing maps for cellular in silico staining and cell
  substate classification.
\newblock \emph{Frontiers in Immunology}, 12:\penalty0 765923, 2021.

\end{thebibliography}
\bibliographystyle{icml2022}

\end{document}